\documentclass{emulateapj}

\newcommand{\sech}{{\rm sech}}

\shorttitle{SIMULATING GALACTIC DISKS}
\shortauthors{TASKER AND BRYAN}

\begin{document}


\title{Simulating Star Formation and Feedback in Galactic Disk Models}

\author{Elizabeth J. Tasker}
\affil{Oxford University, Astrophysics, Keble Road, OX1 3RH, UK}

\and

\author{Greg L. Bryan}
\affil{Department of Astronomy, Columbia University, New York, NY 10027}

\begin{abstract}
We use a high-resolution grid-based hydrodynamics method to simulate the multi-phase interstellar medium in a Milky Way-sized quiescent disk galaxy.  The models are global and three-dimensional, and include a treatment of star formation and feedback.  We examine the formation of gravitational instabilities  and show that a form of the Toomre instability criterion can successfully predict where star formation will occur.  Two common prescriptions for star formation are investigated.  The first is based on cosmological simulations and has a relatively low threshold for star formation, but also enforces a comparatively low efficiency.  The second only permits star formation above a number density of $10^3$ cm$^{-3}$ but adopts a high efficiency.  We show that both methods can reproduce the observed slope of the relationship between star formation and gas surface density (although at too high a rate for our adopted parameters).  A run which includes feedback from type II supernovae is successful at driving gas out of the plane, most of which falls back onto the disk.  This feedback also substantially reduces the star formation rate.  Finally, we examine the density and pressure distribution of the ISM, and show that there is a rough pressure equilibrium in the disk, but with a wide range of pressures at a given location (and even wider for the case including feedback).
\end{abstract}



\section{Introduction}

Star formation in galactic disk systems is the product of a large number of physical processes and so is potentially quite complicated.  In outline, gravity tries to form the dense molecular clouds out of which stars form, while rotational shear, thermal pressure, turbulence, magnetic  fields, and cosmic ray pressure resist the collapse.  Since the gas in disks does not turn into stars on a free-fall time, one or more of these resistive mechanisms must be effective.  The classic condition for disk instability, the Toomre $Q$ parameter \citep{toomre64} encodes the impact of shear, which suppresses large-scale perturbations, and the effective sound speed, or pressure (which suppresses small-scale fluctuation, and may be any of the physical processes described earlier).  When $Q$ is above some critical value, the large and small-scale suppression ranges overlap and no (linear) perturbations are gravitationally unstable.

While this picture is theoretically pleasing and has substantial observational support \citep{kennicutt89,boissier03,heyer04}, it is still not perfectly clear which of the physical processes listed above actually supply the local effective pressure.  Probably the leading candidate is turbulence because it is seen in all disks and has many of the correct properties, but it should be noted that the magnetic fields and cosmic-ray pressure have similar energy densities \citep{boulares90}.  Turbulence in galactic disks can be generated from a number of sources, including gravity \citep{wada02}, stellar winds, supernovae \citep[e.g.,][]{MacLow2004}, the magneto-rotational instability \citep{sellwood99} and even radiative heating \citep{kritsuk02}.  Whatever the source, the combined effect of the effective pressure, gravity and shear must reproduce both the observed threshold density for star formation and also the observed relation between gas surface density and star formation \citep{schmidt59, kennicutt89, kennicutt98, martin01}.

A related question is the structure and distribution of gas densities, temperatures and pressures within the interstellar medium (ISM).  Observational and theoretical work have suggested a picture of a multi-phase medium with substantial turbulent motions \citep{mo77, mccray79, larson81, stanimirovic99, 2000MNRAS.315..479D, elmegreen01}.  However, the distribution of gas (both in terms of volume and density) in the various phases is not well understood, although substantial observational progress has been made \citep{jenkins78, shelton94, ferguson96, chu99, shelton01}.  It is clearly important to try to probe this topic theoretically in order to tease apart the connection between star formation, feedback and the ISM, not simply for a better understanding of our galaxy and local galaxies, but also to model star formation and galaxy formation at high redshift.

Numerical work often focuses on one of two aspects; either a detailed analysis of the ISM and a smaller simulation area, or a study of the global disk instabilities and star formation at the cost of a simplified ISM model. Models that have tackled both topics have either been in  two-dimensions \citep{wn2001} or restricted to a box size a few hundred parsecs across \citep{2001ApJ...559L..41W}.

Work performed in two dimensions by Rosen \& Bregman (1995) allowed the ISM to evolve self consistently, but treated the stars as a collisional rather than collisionless fluid.  They modeled the galaxy side-on so that the simulation region included a dimension out of the disk. They found that the gas formed a three-phase medium with cold and warm filaments surrounding bubbles of hot gas. The bubbles of hot gas extend to up to a kiloparsec across, with filamentary structure similar to that observed in our own Galaxy. 

Self consistent treatment of the ISM and star formation has been performed in two dimensions (with both dimensions in the plane) by \citet{wn2001} and in three dimensions over a small box size \citep{2001ApJ...559L..41W}. They see three phases, but also gas that exists in unstable regions between these phases.  They therefore argue that a simple two or three phase model of the ISM is not sufficient to represent it properly, and that turbulence results in the smearing out of the phases so that gas exists outside pressure equilibrium. Wada also finds that the hot gas is a product of the supernovae explosions and both the hot and warm gas exist off the surface of the disk, leaving the cold gas on the disk plane.

In three dimensions but considering a small section of disk, \citet{2000MNRAS.315..479D} also utilized a separate stellar disk to explore the collective effects of type I and II supernovae on the structure of the ISM. His simulations were performed in three dimensions for a section of a galactic arm, located $8.4$\,kpc from the galactic centre using a fixed gravitational field. The supernovae locations were determined randomly but with constraints imposed to give a realistic distribution. His results show cold gas is present in a thin, irregular layer on the galactic plane, intercrossed with tunnels of hot gas from supernovae explosions. Around the cold gas is a thick disk of neutral warm gas up to $500$\,pc followed by ionized warm gas and then hot gas at heights above $1.5$\,kpc. Places where several supernovae merge form reservoirs of hot gas that have enough energy to break free of the gravitational pull of the stellar disk and expand upwards in large bubbles. 

Korpi \citep{1999ApJ...514L..99K} modeled a section of the ISM self consistently, including  the effects of type I and II supernovae and that of magnetic fields, but left out star formation and self gravity. Their ISM formed a two phase structure of warm and hot gas with a bimodal temperature-density distribution. The warm gas was found in scale heights less than $500$\,pc and the hot gas above that. They also found a cold component, due to compression by the warm gas, which was found at heights of less than $100$\,pc. The supernovae in their simulations clustered to produce large non-spherical shells. 

Several numerical studies have simplified their treatment of the ISM to study the global properties of the disk and star formation.  Robertson et al. \citep{2004ApJ...606...32R} and Semelin \& Combes (2002) both assume a two-phase ISM consisting of cold clouds embedded in a warm gas in pressure equilibrium. The ratio of gas in these two phases is controlled statistically by allowing gas to switch phases during supernovae explosions, conduction and cooling processes. 

In their paper, \citet{2004ApJ...606...32R} compared simulations which used first no star formation, then star formation but no feedback and finally feedback with a two phase ISM. In the non-feedback cases, the gas cooled extremely efficiently, resulting in a near isothermal ISM. They found that in these cases, disk fragmentation was catastrophic and the stars (when present) ended up in two big clumps. The addition of star formation stabilised the disk for slightly longer than in the no star formation case. In both cases, the fragmentation was due to the Toomre instability \citep{toomre64}. The addition of feedback and a multiphase ISM resulted in increased pressure support and a smoother distribution of gas and stars. 

\citet{Li2005a,Li2005b} use an isothermal gas for the ISM and examine star formation for two different temperatures. They find in both cases the gradient of the gas surface density plotted against the surface density of the star formation rate is around 1.5, in good agreement with the Schmidt law and the observations of \citet{kennicutt89}. They also observe a threshold for star formation, where no stars are formed past two radial scale lengths. This is also the point where the Toomre Q parameter drops below one and the disk becomes stable to Toomre instabilities. 

\citet{2003ApJ...590L...1K} performed hydrodynamic simulations in a cosmological context. In this work, the gas is converted into stars on a characteristic gas consumption time scale, rather than on the dynamical time. Kravtsov finds that this can still reproduce the Schmidt law with a gradient of 1.4. The addition of feedback in these simulations results in considerably more hot gas at low densities although the PDF of the gas density remains unchanged.

This paper is the first step in a longer range plan to understand the fundamental processes of star formation and feedback in a galaxy disk. These are the first three-dimensional simulations of a global disk without the need to simplify the structure of the ISM, using a grid-based code which is better abled to resolve the multi-phase medium (except for the small-disk simulations of \citet{2001ApJ...559L..41W}, which are also grid-based). We use this model to investigate local star formation throughout the evolution of the disk, from the early fragmentation of the gas into stars through to the global and local properties of the star formation rate and the effect on the evolution of the interstellar medium. We compare the results for two different models of star formation and with and without the inclusion of stellar feedback from type II supernovae. Ultimately, we hope the better understanding of star formation gained from looking at our isolated disk will act as a guide to cosmological simulations of galaxy formation. For these simulations we use a high-resolution adaptive mesh refinement (AMR) code, which includes a more sophisticated treatment of star formation and feedback as well as a full treatment of self-gravity of the gas rather than the fixed potential that is often used. We concentrate on hydrodynamical effects, ignoring (for the moment) magnetic fields and cosmic ray pressure.

In section~\ref{sec:comp_method}, we describe our computational approach, in section~\ref{sec:structure} we discuss the structural properties of the disk simulations, including the formation of instabilities and the vertical distribution.  In section~\ref{sec:sf_properties}, we discuss how  star formation is related to the surface density and compare this to observations, while section~\ref{sec:ism_properties} contains an analysis of the multi-phase structure of the resulting ISM.

\newpage

\section{Computational Methods}
\label{sec:comp_method}

\subsection{The Code}

To model the galaxy disk, we used the structured adaptive mesh refinement (AMR) grid code, {\it Enzo}, described in \citet{Bryan1997}, \citet{Bryan1999}, \citet{Norman1999}, \citet{Bryan2001} and \citet{OShea2004}. {\it Enzo} is a three-dimensional hydrodynamics code that uses a grid-based scheme for the gas and particles for the stars. A large advantage to using AMR codes over static grids is that individual regions of the simulation box can have different levels of refinement. This significantly reduces computational time by only refining areas that need it. An AMR code works by initially placing a single, uniform grid over the whole simulation box. This is the `parent' or root grid and consists of large grid squares allowing a small number to cover the entire volume. The small number of grid cells allows the average properties of each cell to be calculated quickly. However, the detail this grid describes is minimal; anything smaller than one of these cells goes unnoticed. The code therefore looks at each cell and decides whether further resolution is required. In this work, the decision is made based on whether the baryon density is above a threshold value. If it is, a finer `child' grid is placed inside the parent cell and the properties of each of its grid cells are then computed. The process can then be repeated, with the child grid itself becoming a parent grid and so on until the desired level of resolution is reached. The result is a nested structure of grids, with very fine grids only over the areas that require high levels of resolution.

The AMR technique has been particularly successful in resolving the multi-phase nature of the interstellar medium in the presence of star formation and feedback processes \citep{Slyz2005}. Particle based codes tend to over-mix the hot and cold gas which leads to an over estimate of the radiative losses, although algorithmic improvements have alleviated this problem somewhat (the problem and proposed solutions are discussed in  detail in e.g., \citet{Marri2003} and \cite{springel2003}).  Grid codes suffer less from this problem as the mesh allows distinct boundaries to be more sharply resolved.  This is one of the reasons that we use {\it Enzo} to study the instabilities produced in the disk and the resulting ISM structure.

The disk is modeled in a three-dimensional periodic box of side $1$\,h$^{-1}$\,Mpc. The size of the parent grid is $128^3$ and we proceed down to an additional eight subgrids of refinement which gives us a maximum resolution (i.e. minimum cell size) of about $50$ pc (and include some runs  with resolution down to 25 pc). Once set up, the disk was allowed to evolve over a period of $\Delta z = 0.1 \approx 1.4$\,Gyrs.  The simulations were performed using comoving coordinates, although over the small range of redshift examined here, the impact of the expansion is very slight (for completeness we note that the model adopted is a $\Lambda$CDM universe with $\Omega_m = 0.3$, $\Omega_\Lambda = 0.7$ and H$_0 = 67$\,kms$^{-1}$\,Mpc$^{-1}$).

The gas is evolved using a 3-dimensional version of the ZEUS hydrodynamics algorithm \citep{Stone1992}. Radiative gas cooling follows the cooling curve of \citet{Sarazin1987} down to temperatures of $10^4$\,K. Further cooling down to $T_{\textrm{min}}=300$\,K is introduced in the second  simulation and used thereafter, where the cooling curve is extended using rates given in \citet{Rosen1995}.  This is larger than the minimum temperature of dense molecular clouds but is in the upper range of temperature for the cold neutral medium \citep{Wolfire2003}.  In their paper, \citet{Rosen1995} argue that other physical processes (e.g. Cosmic-ray pressure or magnetic fields) may be crudely modeled by such a choice.  More practically, it allows us to observe the formation of a multi-phase medium including the primary phases commonly discussed and yet resolve the Jeans length over most of the disk (we will return to this issue in more detail).

For the simulations in which star formation was allowed to occur, the following criteria were used to decide whether a grid cell would produce a star \citep{Cen1992, OShea2004}: (i) the gas density in that grid cell exceeds a threshold density. (ii) the mass of gas in the cell exceeds the local Jeans mass, (iii) there is convergent flow (i.e. $\nabla \cdot v < 0$) and (iv) the cooling time is less than the dynamical  ($\tau_{\textrm{cool}}<\tau_{\textrm{dyn}}$), or the gas temperature is at the minimum allowed value. If a grid cell meets all the previous criteria then some gas is converted into a `star particle'. The mass of this star particle is calculated as:
\begin{equation}
m_{*}=\epsilon\frac{\Delta t}{t_{\textrm{dyn}}}\rho_{\textrm{gas}}\Delta x^3
\label{eq:sfr}
\end{equation}
where $\epsilon$ is the star formation efficiency (more properly the efficiency per dynamical time), $\Delta t$ is the size of the time step, $t_{\textrm{dyn}}$ is the time for dynamical collapse and  $\rho_{\textrm{gas}}$ is the gas density.  This set of conditions has one extra criteria added to it: even if a cell fulfills all of the previous four criteria, a star particle will not be formed if its mass would be less than a minimum star particle mass $m_{\textrm{*min}}$. In most of our simulations, the value for $m_{\textrm{*min}}$ used was $10^5$\,M$_\odot$. The reason for this addition is purely computational: a large number of small stars would greatly slow down the simulation. However, in the case where this criteria is the only mechanism preventing a star particle from forming, a bypass exists that allows a star particle with mass less than $m_{\textrm{*min}}$ to form with a probability equal to the ratio of the mass of the predicted star particle over $m_{\textrm{*min}}$ (in which case the resulting star mass is $m_{\textrm{*min}}$ or 80\% of the mass in the cell, whichever is smaller).

\begin{figure*} 
\begin{center} 
\includegraphics[width=15cm]{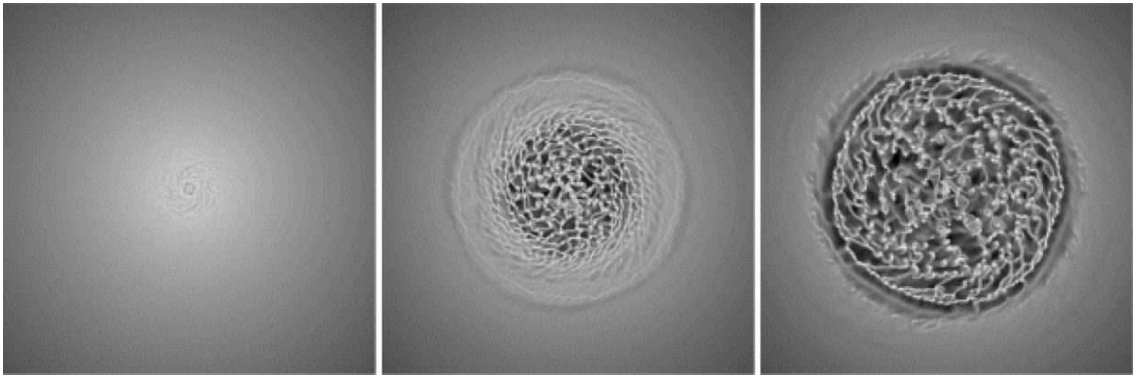}
\caption{The evolution of the gas density as a function of time for the DHIRES run in the central 21.4 kpc of the run.  From left to right, the frames are at 50 Myr, 100 Myr and 150 Myr, respectively.  The density ranges from $10^{-1}$ to $10^3$ M$_\odot$ pc$^{-2}$.
\label{fig:growth}}
\end{center} 
\end{figure*}

\begin{table}
\caption{Simulation Parameters}
\begin{tabular}{cccccc}
& T$_{\rm min}$ (K) & min $\Delta x$ (pc) & $\epsilon$ & $n_{\rm thresh}$ (cm$^{-3}$) & Feedback \\
\tableline
A & $10^4$ K       & 50 & 0      & 0.02  & No\\
B & $10^{2.5}$ K & 50 & 0      & 0.02 & No\\
C & $10^{2.5}$ K & 50 & 0.05 & 0.02  & No\\
CHIRES & $10^{2.5}$ K & 25 & 0.05 & 0.02  & No\\
CLOWEFF & $10^{2.5}$ K & 50 & 0.005 & 0.02  & No\\
CFDBCK& $10^{2.5}$ K & 50 & 0.05 & 0.02 & Yes\\
D & $10^{2.5}$ K & 50 & 0.5 & $10^3$  & No\\
DHIRES & $10^{2.5}$ K & 25 & 0.5 & $10^3$  & No\\
DFDBCK & $10^{2.5}$ K & 50 & 0.5 & $10^3$ & Yes \\
DJEANS & $10^{2.5}$ K & 25 & 0.5 & $10^3$  & No\\
DCONST & $10^{2.5}$ K & 50 & 0.5 & $10^3$  & No\\
\end{tabular}
\end{table}

To model the star formation in a molecular cloud, which will typically spread out over a dynamical time, the star particle's formation is spread out over time such that its mass at a time t is:

\begin{equation}
m_{\textrm{stars}} (t) = m_*\int^{t}_{t_{SF}}\frac{(t-t_{\textrm{SF}})}{\tau^2}\exp\frac{-(t-t_{\textrm{SF}})}{\tau}dt
\end{equation}
where $m_*$ is the mass of the star particle, $t_{\textrm{SF}}$ is the time the star particle was formed and $\tau = max(t_{\textrm{dyn}},10$\,Myr).

The code also allows the inclusion of stellar feedback from type II supernovae explosions. This form of feedback has often been suggested as the main driving force for self-regulated star formation. If this option is switched on in the code (as it is for two of our simulations), then $10^{-5}$  of the rest-mass energy of generated stars is added to the gas' thermal energy over a time period equal to $t_{\textrm{dyn}}$. This is equivalent to a supernova of $10^{51}$ erg for every $55$\,M$_\odot$ of stars formed. All this energy goes into the cell in which the star particle has been created. 


\subsection{Initial Conditions}

Our simulations start with an isothermal gas disk with a temperature of $10^4$ K and a density profile given by: 
\begin{equation}
\rho(r,z) = \rho_0 e^{-r/r_0}\sech^2\left(\frac{1}{2}\frac{z}{z_0}\right).
\end{equation}

Integrating this expression for the density gives us the total mass $M_{\textrm{gas}}(r) = 8\pi\rho_0 R_0^2 z_0$, where $\rho_0 = 2.36\times 10^{-20}$ kgm$^{-3}$ which is based on a total gas mass of $1\times 10^{10}$\,M$_\odot$.  We note that this value is low compared to the Milky Way  gas disk, which is roughly 4 times higher \citep[e.g.,][]{klypin1997}. For the scale radius and height we took typical values of $r_0 = 3.5$ kpc and $z_0 = 400$ pc respectively.

The disk sits in a dark matter profile which takes the form described by \citet{navarro97}. This produces a dark matter mass at a radius $R$ of:
\begin{equation}
M_{\textrm{DM}}(r)=\frac{M_{200}}{f(c)}\left[\ln(1+x)-\frac{x}{1+x}\right],
\end{equation}
where the virial mass,  $M_{200} = 10^{12}$\,M$_\odot$, $x = Rc/r_{200}$ and the concentration parameter $c = 12$. $f(c)$ is an expression given by:
\begin{equation}
f(c)=\ln(1+c)-\frac{c}{1+c}
\end{equation}

Adding together the gaseous and dark matter mass components allows us to calculate the initial circular velocity of the disk using $V_{\textrm{circ}}(R) = \sqrt{GM_{\textrm{tot}}/R}$.

\subsection{Summary of the Performed Runs}

Table 1 presents the simulations we ran, outlining the different parameters used in each run.  There are four main groups of simulations, the first (A) permitted cooling only down to $10^4$ K, which is the minimum allowed by neutral hydrogen line cooling.  Simulation B (and the remainder of the  simulations) was allowed to cool down to our lower limit of 300 K.  Simulation C included star formation with an efficiency typical of large-scale cosmological simulations (5\%).  This efficiency is appropriate for the galactic disk as a whole, so we adopted a low density threshold which allows stars to form even in relatively low density regions (provided they pass the other criteria).  In this picture, we admit we are not following the formation of all dense clumps and so use a Schmidt-like law to model the star formation rate.  Simulation D on the other hand, assumes that stars will only form in the giant (molecular) clouds resolved in the simulation and so we adopt a high efficiency and a high threshold (corresponding approximately to a number density of 10$^3$ cm$^{-3}$).  In addition to the basic C and D runs, we also perform a number of variants.  In each case, we include high resolution runs (CHIRES and DHIRES) which use a root grid with twice as many cells and so twice the spatial resolution and 8 times the mass resolution of the standard runs and also runs which include feedback from type II supernovae (CFDBCK and DFDBCK).  In addition, we examine the impact of reducing the star formation efficiency by a factor of ten (CLOWEFF) and adding an additional refinement criteria that forces the local jeans length to be resolved by at least four cells (DJEANS), as suggested by \citet{Truelove1997} (at least until we reach the maximum refinement level, or the minimum $\Delta x$ given in Table~1). We also perform one final simulation (DCONST) where we changed our function for star particle mass as given in (\ref{eq:sfr}) to depend on a set time-scale, rather than on the dynamical time.


\section{Structural Results}
\label{sec:structure}

In this section, we investigate how the disk cools and fragments and the impact that star formation has on the three-dimensional structure.

\begin{figure*} 
\begin{center} 
\includegraphics[width=15cm]{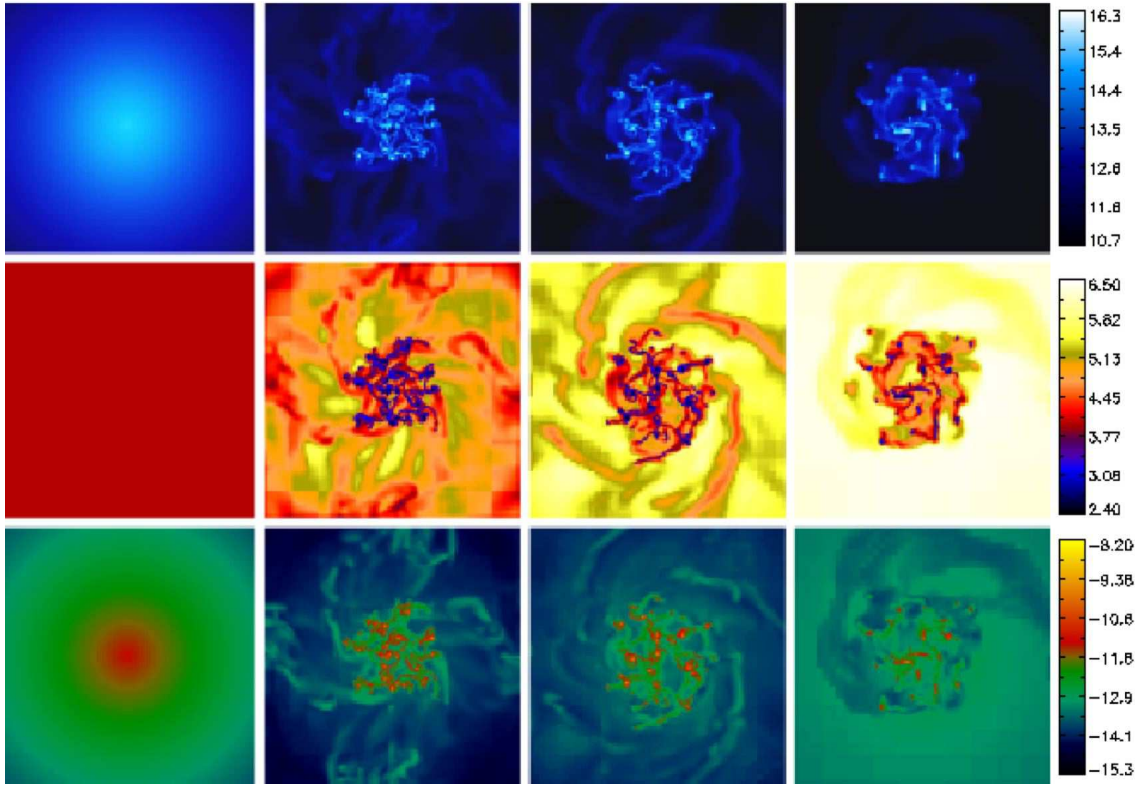}
\caption{Gas surface density (top), mean mass-weighted temperature (middle) and mean pressure (bottom) in the inner $30$\,kpc region of the simulation box for run B.    From left to right, the frames show the results at: t = 0, 374 Myr, 561 Myr and 1.32 Gyr.  All scales are to the $\log_{10}$ and gas and star particle density is measured in M$_\odot$ Mpc$^{-2}$, temperatures in K and pressure on an arbitrary scale.
\label{fig:run2}}
\end{center} 
\end{figure*}

\subsection{Evolution and Structure of the Disk}

In the first simulation performed (simulation A in Table~1, with cooling limited to a minimum temperature of $10^4$ K), the disk remained largely uniform with only small fluctuations occurring in the central few kpc.  As we will show below, this is consistent with our expectations from the  growth of gravitational instabilities.  No star formation would have occurred even if we had allowed it.

However, when cooling is allowed down to 300 K (as in runs B, C and D), the result is quite different.  Figure~\ref{fig:growth} shows the growth of the perturbations over the first 150 Myr.  The perturbations start in the center where the dynamical time is the shortest (recalling that $t_{dyn}  \sim \rho^{-0.5}$) and spread outward with time.  Initially, the perturbation begins as a spiral density wave as the radial direction collapses first.  This forms spiral filaments which then fragment in the azimuthal direction.  Knots quickly appear and accrete matter along the filament, in a manner reminiscent of cosmological structure formation.  Eventually the filament disappears and neighbouring knots start to interact.  The interactions lead to mergers but also to high-velocity encounters which can strip material from the knots.  Eventually the entire face of the disk which is prone to instabilities develops these clumps.  The presence or absence of star formation in our simulations does not significantly change this picture.

These images are similar in many respects to the 2D simulations presented in \citet{wn2001}, although on a somewhat larger scale.  Increasing the dimensionality does not appear to significantly change the early evolution of the clumps.  However, examining images of disk in a  side-on projection shows that the clumps do make excursions out of the plane during interactions and the later evolution we see differs significantly from the 2D Wada \& Norman results.  We will discuss the vertical structure in more detail below.

Figure~\ref{fig:run2} shows the later evolution of run B, with cooling. The projections here show the evolution of the disk pressure (bottom) in addition to the gas density (top) and temperature (middle) for four of the simulations outputs. The dense knots show up clearly in the temperature  plot as the coldest gas.  This is consistent with the very short cooling times implied by such dense gas.  Basically, the gas is as cold as our truncated cooling curve allows.  The anti-correlation between density and temperature implies some degree of pressure balance.  However, the bottom set of panels in this figure show that while this is true for the low and moderate density gas, the highest density clumps are significantly over-pressured compared to the rest of the ISM (note that the pressure here is only thermal pressure - we do not try to characterize any turbulent component to the pressure). Further out in the disk (beyond about 10 kpc), the gas fails to form cold, dense clumps.  The gas also shocks up to temperatures around $10^5 - 10^6$ K (also the temperature of the gas in the inter-clump regions near the center of the disk).

In runs C and D, star formation is introduced via the parameters described in the previous section.  The initial evolution is quite similar in that the gas quickly fragments, but now the addition of star formation depletes gas from the self-gravitating clumps. Stars only form in  the central 10 kpc (where cold clumps exist in figure~\ref{fig:run2}).  On a time-scale of several hundred Myr, the gas is largely converted to stars and the gas density of the disk drops below that required to sustain instabilities, cutting off star formation except in the innermost regions.  


\subsection{Disk Instability}
\label{sec:instability}

\begin{figure} 
\begin{center} 
\includegraphics[width=8.5cm]{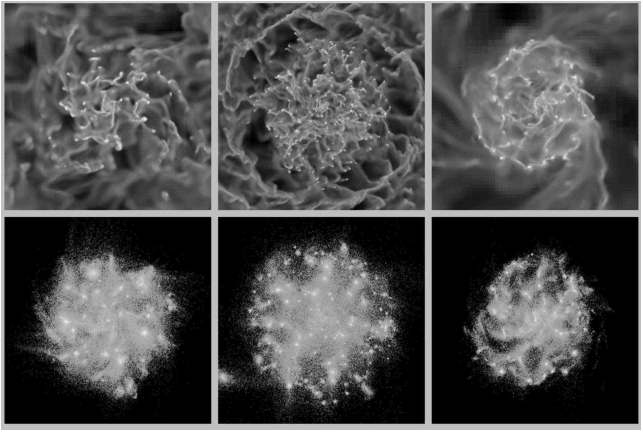}
\caption{This figure shows the gas density (top) and stellar density (bottom) in a region 22 kpc on a side after 330 Myr of evolution.  The leftmost images are from the simulation with star formation but no feedback; the central images are from the high-resolution version of this  simulation, while the right-most images are from the feedback simulation.
\label{fig:dens_star_6panel}}
\end{center} 
\end{figure}

In figure~\ref{fig:dens_star_6panel}, we show the gas and stellar densities of the simulations with and without feedback, including a non-feedback star formation run at high-resolution.  It is clear that the typical clump size is strongly affected by resolution, with smaller fragments  appearing at higher resolution.  This is consistent with the fact that we do not always resolve the Jeans length in the center of the disk, particularly for the standard D run.  On the other hand, there is a well-defined radius beyond which star formation does not occur, and this does appear to be well-resolved.  In particular, a run for which we ensure that the Jean's length is resolved (DJEANS) up to the maximum resolution  of the adaptive mesh, produces results which are essentially identical.  For gas at our minimum temperature (300 K), the Jean's length is 25 pc at a number density of $10^2$ cm$^{-3}$; thus, in the DHIRES run we resolve the Jeans length nearly until the threshold at which gas is converted into stars (10$^3$ cm$^{-3}$).  Therefore, we believe that the cutoff in star formation beyond a radius of 7 or 8 kpc seen in figure~\ref{fig:dens_star_6panel} is a robust result (this low radius compared to the Milky Way is due to our assumed low disk density - a more realistic gas disk would have a larger cutoff radius).

\begin{figure} 
\begin{center} 
\includegraphics[width=7cm,angle=0]{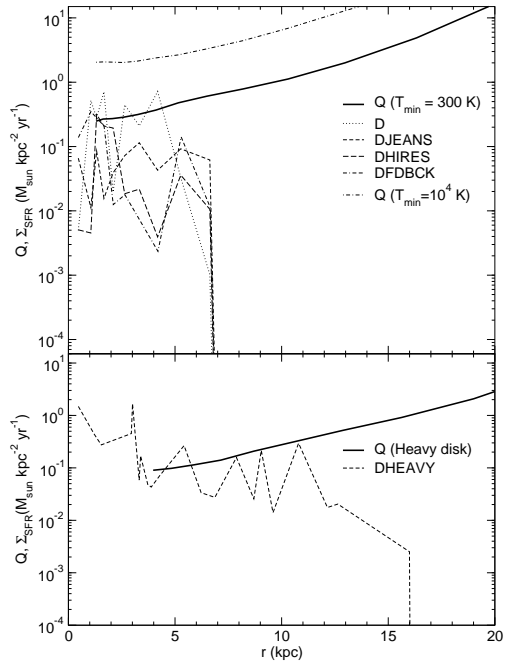}
\caption{In the top panel, the solid line shows the $Q$ parameter computed shortly after the start of the D simulation when the gas has reached its minimum value (300 K) but before non-linear instability formation.  The instantaneous rate of star formation (averaged over 20 Myr) is shown for four variants of the  D simulation (the C series shows similar results).  Note that gravitational instability (and hence star formation) only occurs below a critical Toomre $Q$ parameter.  The dot-dashed curve shows the $Q$ parameter for the A simulation ($T_{\rm min} = 10^4$ K) which never falls below this critical value. The bottom panel shows the same plot for a disk with a gas mass four times greater than the other simulations. The cutoff this time it at a larger radius and corresponds to a slightly larger $Q$ value. 
\label{fig:toomreQ_sfr}}
\end{center} 
\end{figure}

This star formation threshold can be understood with the Toomre stability parameter \citep{toomre64}, defined as $Q = \kappa c_s / \pi G \Sigma_g$, where $\kappa$ is the usual epicyclic frequency, $c_s$ is the thermal sound speed as measured in the disk (about 2 kms$^{-1}$ for our minimum  temperature of 300 K), and $\Sigma_g$ is the gas surface density.  This parameter is plotted as a function of radius in figure~\ref{fig:toomreQ_sfr}, along with the current star formation rate (averaged over the last 20 Myr) at that radius for a variety of simulations.  In each case, there is a sharp cutoff at a particular value of the $Q$ parameter. From the runs represented in the top panel of figure~\ref{fig:toomreQ_sfr}, we see this cutoff is unaffected by feedback. As we describe below, feedback acts to reduce the star formation rate, but it has no baring on this stability cutoff point. The critical $Q$ parameter derived this way is about 0.6.  A linear analysis for a 2D dimensional disk predicts 1 \citep{toomre64}, while a finite disk thickness reduces the critical value to 0.676 \citep{goldreich65, gammie01}.  We note that \citet{kennicutt89} finds values around 1.5, however we have used the thermal sound speed of the gas rather than the velocity dispersion.  If we adopt an effective sound speed of 6 kms$^{-1}$, this boosts the derived critical value by a factor of 3 and brings it into better agreement with the Kennicutt value. 

For comparison, we also calculated the star formation rate and $Q$ for a disk with four times the gas mass which is shown in the bottom panel of figure~\ref{fig:toomreQ_sfr}. The heavier disk draws out the star formation cutoff to a greater radius, in close agreement with the Milky Way's own stellar radius of $\sim 15$\,kpc. The value for $Q$ at the absolute cutoff for the star formation, $r=16$\,kpc, is 1, a slightly higher than for the original disk. However, the star formation rate starts to decrease at a smaller radius than this, at around $r=13$\,kpc. At this point, $Q\sim 0.5$, a closer agreement to the lighter disk. The $Q$ scaling therefore works well with the changing weight of the disk, especially since the variation of $Q$ over the disk is of order 100.  This scaling could be tightened still more with a sharper estimate of the correct position of the star formation rate cutoff.


\subsection{Vertical Scale Height and the Galactic Fountain}

\begin{figure} 
\begin{center} 
\includegraphics[width=7cm]{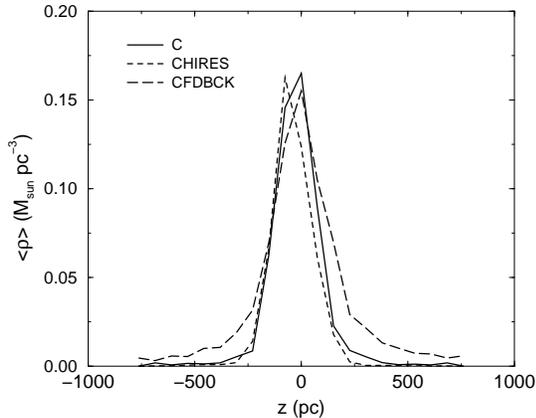}
\caption{The vertical density profile 330 Myr into the evolution for the same three simulations shown in figure~\ref{fig:dens_star_6panel}.  This is the density profile averaged within two radial scale-heights.
\label{fig:height_profile}}
\end{center} 
\end{figure}

Another important structural property is the vertical scale-height.  In figure~\ref{fig:height_profile}, we show the density profile for the same three simulations discussed previously.  Both the low and high-resolution simulations without feedback show a thin disk, with a scale-height of  approximately 100 pc.  This does not change significantly for the feedback simulation, although there is a clear tail of higher density material which extends about a kpc above and below the disk.  This is due to the transient gas streams which are ejected from the disk due to supernovae explosions.  This scale-height is comparable but less than the observed Milky Way HI scale-height which ranges from about 150 pc at a radius of 5 kpc, to 300 pc at the solar radius \citep{Malhotra95}.  Unfortunately, even our highest resolution simulation has a cell-size of only 25 pc, so this is, at best, marginally resolved.

\begin{figure} 
\begin{center} 
\includegraphics[width=7cm]{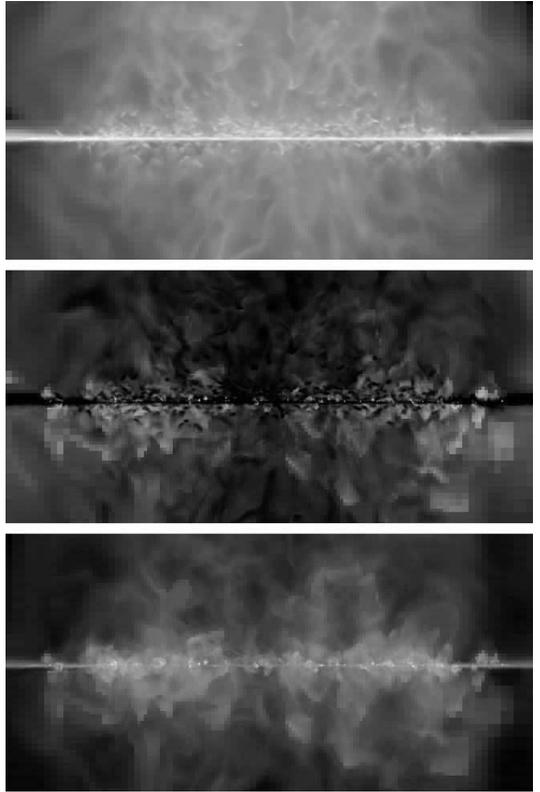}
\caption{This figure shows edge on projections of the CFDBCK run demonstrating the impact of SN feedback (runs without feedback show a thin, cold disk with hot, featureless gas above and below the plane).  The top panel shows the projected density, the center show the mean temperature, while the bottom panel depicts the integrated pressure.  Each image is about 22 kpc across and 10 kpc high.
\label{fig:side_disk}}
\end{center} 
\end{figure}

This is not to say that the feedback run doesn't have a substantial impact on the distribution of gas in the plane.  Figure~\ref{fig:side_disk} shows side-on projections of density, temperature and pressure for the CFDBCK run.  While it is clear from these three plots that most of the gas still lies in the plane, large streamers can be seen reaching for kpc above and below the plane.  Hot bubbles can be seen in the temperature distribution with some gas around $10^6$\,K, but most considerably cooler.  These bubbles form in the plane but many quickly break out, driving warm, diffuse gas upwards.  As this gas expands and cools it is also de-accelerated by halo gas.  Clumps of the gas cool and fall back towards the plane.  Many of these can be seen at dark spots in the temperature distribution but cannot be seen in the pressure image, indicating that they are in a rough pressure equilibrium with the hotter, more diffuse medium.  Although this galactic fountain effect does not significantly change the density profile near the plane, it does very clearly change the dynamics of gas out of the plane.  We will examine the observational consequences in a future paper.

\section{Star Formation Properties}
\label{sec:sf_properties}

Observational studies of star formation indicate that the rate of star formation is closely tied (in a statistical sense) to the surface density of the gas, increasing as $\Sigma_{\rm SFR} \propto \Sigma_{\rm gas}^{1.5}$ \citep{kennicutt89}.  In this section, we examine how star formation  occurs in our model galaxies.

\subsection{Star Formation History}

\begin{figure} 
\begin{center} 
\includegraphics[width=8cm]{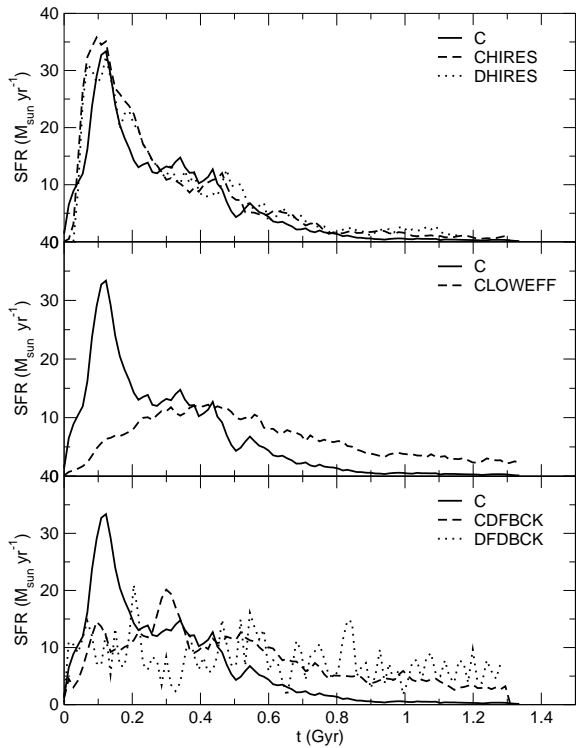}
\caption{The star formation rate as a function of time for the models indicated in the legend.
\label{fig:sfr}}
\end{center} 
\end{figure}

First, we examine the star formation history in the entire disk as a function of time, as shown in figure~\ref{fig:sfr}.  The top panel depicts the results for our model C, both the standard run and the high resolution, CHIRES, variant along with the high resolution run for the D-algorithm, DHIRES. The close agreement between these  curves again demonstrates that the star formation is largely resolved in these simulations.  Each curve shows a rapid increase in the star formation rate over roughly 100 Myr, in agreement with the images in figure~\ref{fig:growth}.  The rate then reaches a peak and falls off in an exponential fashion.  This is due to the gas depletion resulting from the high star formation rate.  After 1 Gyr, nearly all of the gas in the unstable part of the disk is exhausted.  This is considerably shorter than the time-scale for gas exhaustion in present day spirals and is related to the higher than observed star formation that we see in these simulations, a point which we will return to in the next section.

Remarkably, the star formation history is nearly identical for the D-model for star formation, which is more efficient but limited to the densest parts of the clumps. In fact this agreement is  largely due to our choice of parameters.  This is shown clearly in the middle panel of the same figure for the CLOWEFF simulation, which is similar to the C run but with 10 times lower efficiency.  This lower efficiency run has a much longer time-scale for converting gas into stars, with a lower peak rate and more gas left after 1 Gyr.  This demonstrates the importance of parameter selection for this type of star formation algorithm (which is very common in cosmological simulations).  However, note that the time-scale for gas-exhaustion is a factor of about 4 longer in the CLOWEFF run compared to the C run, not the factor of 10 one would expect by decreasing the efficiency by a factor of 10. This occurs because the longer time-scale permits larger mass clumps to form (mostly through merging), and the larger clouds generate larger core densities, which increases the star formation rate.

In contrast with the C-type algorithm, the parameter choice (within reasonably bounds) is not important in our D type run.  In a test run with the D-type parameters but with 10 times lower efficiency, the result was nearly identical.  This can be readily understood - star formation is only  permitted in dense regions where the dynamical time is so short that the gas will be converted into stars in a time short compared to the galactic time-scale (the dynamical time for gas at a number density of $10^3$ cm$^{-3}$ is about a million years, so even increasing this by a factor of 10 by decreasing the efficiency by 10 would have little effect when compared to the 100 Myr timescales of the disk itself).

Finally, the simulations which includes feedback from type II supernovae with the C-type (CFDBCK) and D-type (DFDBCK) algorithms are shown in the bottom panel. These both show a longer time-scale for star formation and like CHIRES and DHIRES, depict very similar histories.  The extended time-scale for star formation is in part because individual star forming clumps are dispersed by the feedback before they can be converted entirely into stars, and in part because  winds from star forming clusters disperse and heat gas in nearby clouds before they can form.  The net result is that star formation is (at least partially) self-regulated.


\subsection{Star formation as a function of gas surface density}
\label{sec:sf}

\begin{figure} 
\begin{center} 
\includegraphics[width=8cm]{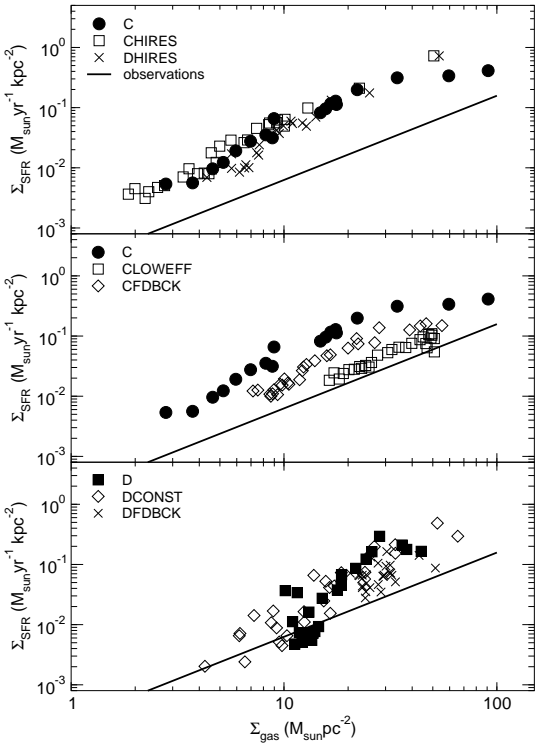}
\caption{The global Schmidt law: variation over time of the relationship between total star formation rate and gas surface density. Star formation rate is plotted as a function of density averaged over the disk for a variety of models.  Each point with the same symbol represents the same simulation at a different time, equally spaced along the 1.4 Gyr simulation run.  The solid line is a best fit from observations \citep{kennicutt89}.
\label{fig:global_sf}}
\end{center} 
\end{figure}

Next, we examine how star formation depends on density, as observations indicate it does.  In figure~\ref{fig:global_sf}, we show the global star formation rate averaged over the star forming part of the disk (corresponding to the radius containing 95\% of the star formation).  The top two  panels show results for the same set of simulations as in figure~\ref{fig:sfr}.  The C and CHIRES runs are quite similar and run nearly parallel to the observed relation between star formation and surface density, albeit at a rate almost an order of magnitude above that observed.  This large star formation is in agreement with the short gas exhaustion times already discussed.  Decreasing the efficiency parameter as in the CLOWEFF run produces results which are closer to those observed, although we see again that the decrease in the star formation rate is a factor of 4 rather than the expected order of magnitude decline (we decrease $\epsilon$ from 0.05 to 0.005), for the reasons noted earlier.  The CFDBCK and DFDBCK run also decreases the star formation rate. In the CFDBCK case this can be seen most clearly to be a factor of two reduction for the same star formation efficiency.

One possible reason for the higher than observed star formation rates is related to our inability to model the physics and structure of molecular clouds.  Such clouds are observed to form stars with an efficiency that ranges from 5\% to 30\% \citep[e.g.][]{Lada2003}, depending on if we are  talking about the entire molecular cloud or embedded clusters.  The efficiency parameter $\epsilon$ is really the star formation efficiency per dynamical time (as an inspection of eq.~(\ref{eq:sfr}) will confirm) and if a cloud is permitted to evolve for many dynamical time without disruption then the final fraction of gas converted into stars can be considerably higher (approaching unity).  This is true for most of our simulations since we do not include stellar winds and ionizing radiation which are thought to be a prime reason for the relatively low efficiency (the exception is for the feedback case in which supernovae can disrupt a cloud after about 10 Myr).  Therefore we should bring down our estimate of the star formation rate by a factor of 3-10 (less for the feedback case).

\begin{figure} 
\begin{center} 
\includegraphics[width=8cm]{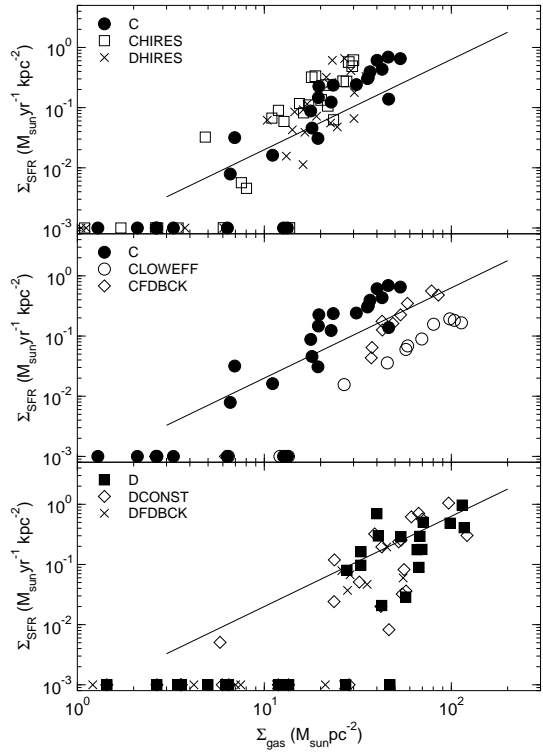}
\caption{The local Schmidt law: variation over the disk's surface of the relation between star formation rate with gas surface density at a set time, t. Star formation rate is plotted against gas surface density averaged azimuthally in radial bins.  Each point (for a given simulation) is averaged over a different radial range, but are all from the same point in time approximately 200 Myr after the start of each simulation (other output  times follow the same relation).  The models shown are the same as in figure~\ref{fig:global_sf}.  Radial bins with no star formation at all are shown at the bottom of each graph.  The solid line is a curve with slope 1.5, as observations indicate (but with an arbitrary normalization).  Note that there is a relatively sharp cutoff  for each set of curves below which there is no star formation, in agreement with figure~\ref{fig:toomreQ_sfr}.
\label{fig:radial_sf}}
\end{center} 
\end{figure}

In figure~\ref{fig:radial_sf}, we show the relationship between star formation rate and surface density for azimuthally-averaged radial bins at a given point in time (unlike the global relation which shows many outputs).  Again, the slope is 1.5, in good agreement with observations, showing  that the simulation reproduce both the global and local relations (or at least their slopes).  Most of the same comments for the top two panels of the previous figure apply to the top two panels of this figure.  Note that in the local relations, there is a relatively sharp cutoff at low disk surface densities below which there is no star formation, as discussed in section~\ref{sec:instability}.

The slope of the simulated relations in both of these plots is very similar to that observed (although the global relations are marginally steeper, particularly at low disk density rates).  For the C simulation series, this is not surprising as this behaviour is built into the star  formation rate in equation~(\ref{eq:sfr}) (the dynamical time is proportional to $\rho^{-1/2}$ so  $\dot{\rho}_{\rm sfr} \propto \rho^{1.5}$).  The D series also uses this relation (but with a higher threshold) and so it appears that here too we have just got out what we put in.  However, this is not the case.  As discussed earlier, the dense clumps are transformed into stars with a net efficiency which is quite high and largely independent of the parameters chosen.  This can be seen in two ways.  First, a simulation with an efficiency ten times lower than the standard D case produced essentially identical results to that shown for the D run.  More convincingly, we have performed a run in which we use a constant time-scale for formation rather than $t_{dyn}$ in equation~(\ref{eq:sfr}), which we denote DCONST (the time constant was chosen to be the dynamical time at the fixed threshold density).  In this case, the local instantaneous star formation rate is directly proportional to the local gas density, however the bottom panels of figure~\ref{fig:global_sf} and \ref{fig:radial_sf} show a very similar scaling as to the standard D run.


\section{Properties of the Interstellar Medium}
\label{sec:ism_properties}

\begin{figure*} 
\begin{center}
\includegraphics[width=15cm]{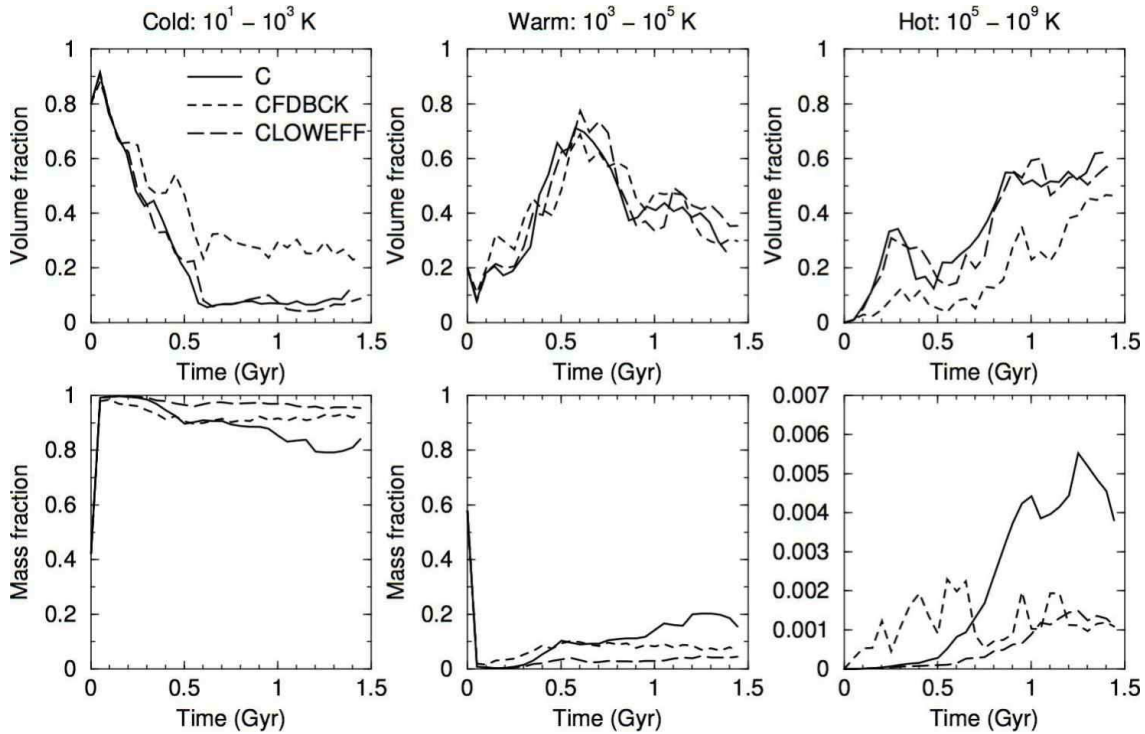}
\caption{{\it top panels:} Fraction of the ISM volume which is cold ($T < 10^3$ K; left panels), warm ($10^3 < T < 10^5$ K; middle panels) and hot ($T > 10^5$ K; right panels), as a function of time for three simulations. Solid lines represents the standard C run, the dotted line is for the run with feedback (CFDBCK), while the long dashed lines show the non-feedback run with a low efficiency of 0.005 (CLOWEFF).
\label{fig:temp_phase}}
\end{center}
\end{figure*}

Previous theoretical studies (McKee \& Ostriker, 1977) have modeled the ISM as a three-phase structure regulated by supernovae explosions. This motivates us to divide the simulated gas up into three temperature ranges: The cold ISM with $T <10^3$ K, warm ISM having $10^3 < T < 10^5$ and  the hot ISM,  $T >10^5$ K. Figure \ref{fig:temp_phase} (top panels) shows the volume of gas in each phase for three of our simulations, while bottom panels show the same results but for the mass fraction.  These have been computed only for gas within 3 radial scale-heights and 400 pc above and below the plane in order to focus on gas in the main disk.

The first order result from these figures is that the volume and mass fractions are generally quite robust to the physical model we use.  Generally, most of the volume is taken by the warm and hot phases, with the cold phase occupying a minority of the space in the disk, except during the first few hundred million years of evolution.   There are some variations about this picture, of course.  In particular, the feedback run (CFDBCK) shows (after about 500 Myr of evolution) a slightly higher volume fraction in the cold phase and a corresponding decrease in the hot phase.  The same panel in the mass fraction plots shows that the mass fraction in this phase is relatively unaffected (and quite high), indicating that the material in this cold phase must be less compressed in the feedback run, probably due to supernova-driven cloud disruption.

The  relatively large fraction of volume in the hot ISM phase, even for the non-feedback runs is somewhat surprising.  This gas is heated by the shocks induced via gravitational instabilities which drive non-circular motions, as well as gas falling from above the plane. The mass fraction in this hot phase is extremely small (note that the scale on this panel differs from the others), indicating that this high temperature material has very low density (and is, in fact, in temperature equilibrium with the other phases).

The mass-weighted phase plots show that most of the mass is in cold, dense clumps (this is particularly true for the CLOWEFF, the low-efficiency case), with the vast majority of the rest in the warm phase.  Only a very small fraction of the mass ever gets heated  significantly; even for the feedback run (CFDBCK), the mass fraction at temperatures above $10^5$ K is always less than one percent.  

We do not find that feedback increases the hot fraction, which is contrary to many models of feedback in the ISM (e.g. McKee \& Ostriker 1977), although the actual fraction of the volume in the hot fraction is in rough agreement with these models.    A possible reason for this is that the type II supernovae we model deposit their energy primarily in dense clumps, leading to their disruption.  Although the clumps are unbound by the supernovae, they are not heated to high-temperatures and quickly cool back down.  This dispersal of dense clouds results in net increase in the volume occupied by cool clouds for the feedback simulation, as the top-right panel shows.


\subsection{Joint Distribution Function}

\begin{figure*} 
\begin{center} 
\includegraphics[width=15.4cm]{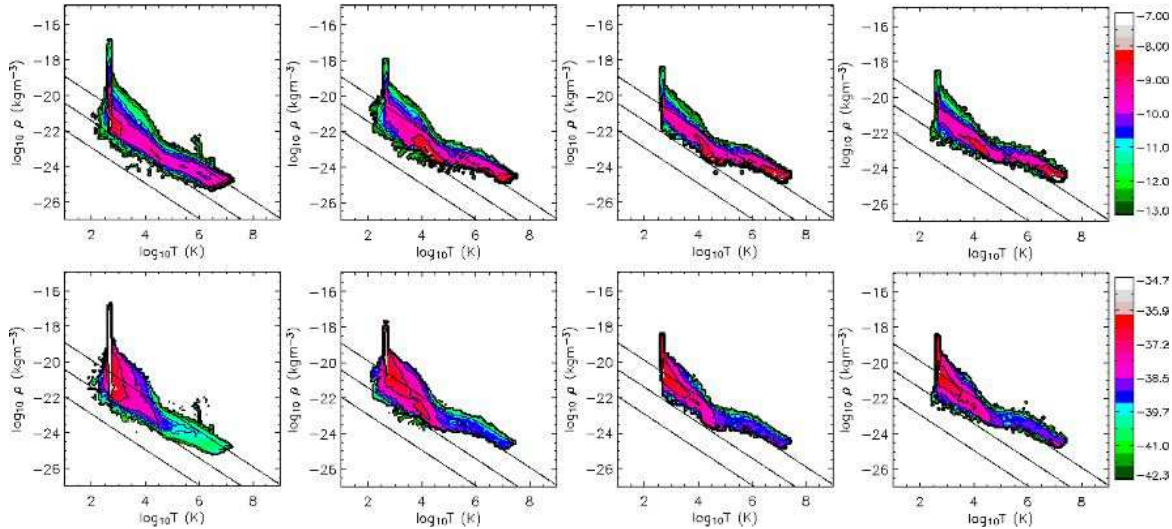}
\caption{Two dimensional contour plots for the volume (top) and mass (bottom) in run C with star formation but no feedback. Plots are shown for $t=191$\,Myr, 567 Myr, 945 Myr and 1.32 Gyr.
Diagonal lines represent lines of constant pressure.
\label{contour_run3}}
\end{center} 
\end{figure*}

A simple three phase model in pressure equilibrium is, of course, an over-simplification. The full nature of the ISM can be better represented in contour plots of density versus temperature for the  volume and mass. Figure \ref{contour_run3} shows the evolution of the gas volume (top) and gas mass (bottom) in the C run for the four outputs at $t$ = 191\,Myr, 567\,Myr, 945\,Myr and $1.32$\,Gyr. The diagonal lines represent lines of constant pressure (assuming an ideal gas equation of state and neglecting ionization changes). The sharp peak on the left hand side of the plots is due to the cutoff temperature (of 300 K) introduced for radiative cooling.

These contour plots make clear that there is not a sharp division between the three phases discussed earlier, but instead a wide distribution of densities, temperatures and pressures.  However, in all of the diagrams weighted by volume, there are peaks at $10^4$ K and $10^{6-7}$ K,  demonstrating that there is some utility in making these divisions.  The gas at $10^4$ K arises because of the sudden drop in the cooling rate at this point, delaying further gas cooling, while the hotter gas comes from shock heating and is (roughly) the virial temperature of the halo.  While there is not an obvious peak at low temperatures in the volume distribution, it is very obvious in the mass distribution, which clearly shows most of the gas residing at the lowest temperature permitted by cooling.  We can see that the contour diagrams support the results from the phase diagrams in that the majority of the gas volume is in the hot and warm ISM phases. The mass contour plot shows the vast majority of the mass is in the cool, dense region and this area shrinks over time as stars are formed from this gas. 

By examining the plots along the diagonal direction of constant pressure we can see first that much of the disk is in rough pressure equilibrium, although the width of the distribution perpendicular to these lines show a spread of at least one order of magnitude.  This indicates that simply  the effect of the gravitational instability can generate a substantial range of pressures.  There are two regions which clearly do not follow the pressure equilibrium.  The most obvious and easiest to explain are the cold, dense clumps which stick up to high densities in a thin line.  These are our self-gravitating clouds.  We remind readers that the position of this line in the temperature direction is set by the cooling cutoff of 300 K; however, it is clear that even if the gas is simply shifted left by 1.5 orders of magnitude (to the minimum observed ISM temperature of 10 K), it would still be over-pressured, even if the density did not increase.  The second feature is the peak at $10^4$ K which actually drops below the line of constant pressure.  This probably is a consequence of the thermal instability which sets in around $10^{5.5}$ K, at which point the gas cools so quickly that it drops out of pressure equilibrium with the surrounding gas \citep[e.g.,][and references therein]{Slyz2005}. 

The evolution in the distribution function over time is relatively moderate, with a mild increase in the pressure from the first to the second frame (over which period the star formation rate drops significantly), and then constant.  This is somewhat surprising considering that most of the gas is converted  to stars during this period.  However, the mass-weighted distribution makes clear that most of the mass is in the dense, self-gravitating clumps and it is this mass (which does not contribute significantly to the pressure in the rest of the plane of the disk) which is converted into stars.

\begin{figure*} 
\begin{center} 
\includegraphics[width=15.4cm]{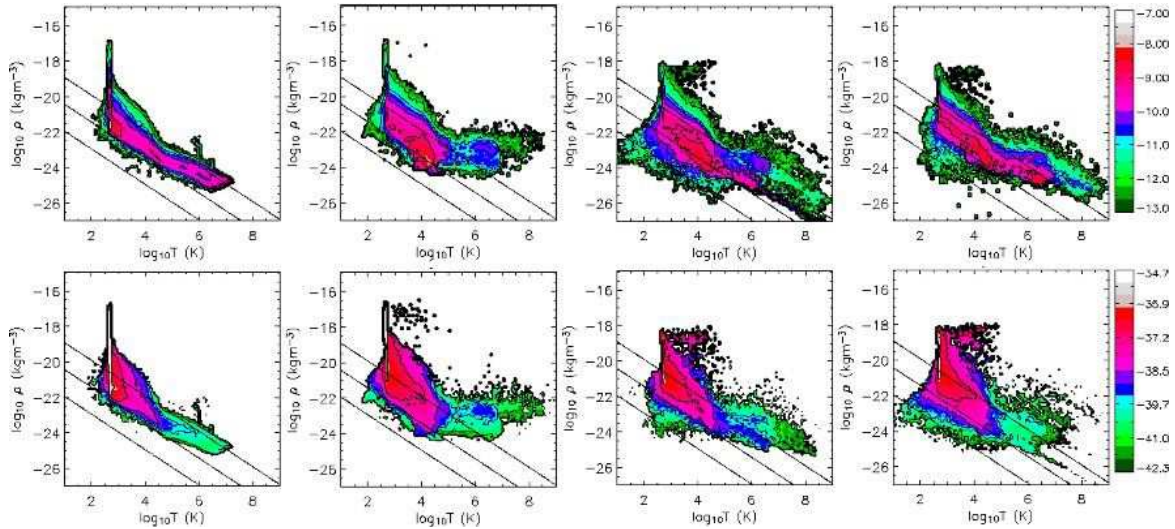}
\caption{Two dimensional contour plots for the volume (top) and mass (bottom) in run CFDBCK with star formation and feedback from supernovae. Plots are shown for the same times as the previous figure.
\label{contour_run4}}
\end{center} 
\end{figure*}

Figure \ref{contour_run4} shows the distribution functions for the run in which feedback is included (CFDBCK). The most striking feature is the dramatic increase in the width of the distribution function. The injection of supernovae energy produces gas with temperatures throughout the full  temperature range to be found at a large range of densities.  The distribution of pressures is also considerably wider than in the non-feedback case, although the median pressure at a given temperature does not appear to be significantly increased.  We also see that the peak in the volume-weighted distribution at high temperatures (and low densities) quickly disperses in the early part of the simulation, but starts to re-emerge in the final image. The cold, high density gas on the far left of the contour plots also seems slower to disperse, in agreement with the slower star formation rate documented earlier. 

\section{Discussion}

What controls star formation in our simulations?  We break this down into two questions, the first is whether stars can form at all at a given radius in the disk, and secondly, at what rate?   As demonstrated in section~\ref{sec:instability}, the growth of instabilities (and hence the  possibility of star formation) is clearly controlled via the Toomre instability criteria.  We derive a critical $Q$ value of approximately 0.6, in reasonable agreement with a linear analysis of finite-thickness disks \citep{goldreich65} and shearing box simulations \citep{Kim2001} that find 0.676 and 0.72, respectively.  Increasing the disk's gas mass by a factor of four, in better agreement with a Milky Way sized galaxy, raises the value of $Q$ to approximately 1, although taking a lower threshold for the star formation cutoff radius gives a $Q$ of around 0.5. $Q$ therefore scales well with the weight of the disk, while refining measurement of the star formation rate cutoff should improve this further. The presence of a stellar disk would modify these results slightly \citep{Kim2001, rafikov2001}.  The addition of a magnetic field with moderate strength can lead to enhanced collapse due to the magneto-Jeans instability and a larger effective $Q$ value \citep{Kim2003}.

Instabilities result in clouds forming over (roughly) a dynamical period and so clearly the rate at which gas can be funneled into the densest regions is set by this timescale.  In the absence of star formation, the clouds are long-lived entities and continue to accrete gas and grow by mergers (in the B  simulation the number of individual clumps drop as the mass of each clump grows until by the end of the simulation only a handful of very large clouds remain).  When star formation is turned on, stars form at the highest rate in the dense centers of these clouds, which tends to create bound clusters; this is particularly pronounced in the D series, where star formation is only allowed in the dense cores.  As long as there is no feedback from the forming stars, their formation does not disrupt the group, although the stars and the gas may become separated due to the different forces on the two components (which is why it is important to model the stars as collisionless particles).  Mergers and further accretion operate on the dynamical time and feed further star formation.  Therefore, it is natural to expect the star formation rate to scale with the local dynamical time as our results demonstrate.

This explains the scaling, but not the amplitude of relation shown in figure~\ref{fig:global_sf}.  As we argued in section~\ref{sec:sf_properties}, this is too large because our effective efficiency for star formation is rather high (near unity integrated over many dynamical times, despite the low $\epsilon$ values adopted), again related to the  longevity of the clusters.  One possible reason behind the overproduction of stars is the lack of photoionization feedback from the newly formed stars.  This will help in a number of ways, as the massive stars will quickly halt star formation in the rest of the cloud which has not collapsed, decreasing the efficiency in each cloud.  It will also lead to the dispersal of the cloud, which will prevent mergers from producing clouds more massive than those observed.  Indeed, in our simulations with type II supernovae, these processes can be seen to operate; leading to a drop by a factor of 2-3 in the star formation rate for a given disk surface density.  The timescale for photoionization is even shorter than the $\sim 10$ Myr period before the first massive stars end their life and explode, which should lead to an even lower net efficiency.

We do not see clear large-scale spiral features in our simulated disks, perhaps indicating that the small-scale perturbations that we observe are insufficient to trigger low-m mode density waves.  It remains to be seen if a more realistic simulation, including satellite perturbers would be  more successful at reproducing grand-design spiral features.

Recently, \citep{Li2005a} have used SPH simulations of quiescent disks to argue that the it is primarily gravity which controls the star formation and gives rise to the Kennicutt relation between star formation and gas surface density.  The results presented in this paper agree in  the sense that gravity, rotation and pressure naturally give rise to the observed slope of the Kennicutt relation, as well as a cutoff in the star formation rate beyond a certain radius.  However, it is equally clear that some sort of feedback from the forming stars is required.  The simulations of \citet{Li2005a, Li2005b} used an isothermal equation of state and argued that an effective sound speed of approximately 10 kms$^{-1}$ would match the observed star formation rates.  They also adopted an efficiency of star formation in molecular clouds of 35\%, again ascribing this to unmodeled feedback effects.  Our simulations with supernovae feedback do indeed reduce the star formation rate, and it is plausible that the addition of photoionization feedback would decrease this further.  It is clear that further work is required.

One of the advantages of a more realistic equation of state is that we can naturally produce a multi-phase ISM, with hot and cold phases existing in rough pressure equilibrium.  A more detailed comparison with observations would be interesting, although a better heating  and cooling model is probably required.  In particular, we do not include a photo-heating source, which probably leads to an over estimate of the amount of cold, dense gas which can cool.  Heating may reduce the star formation rate (although heating will be ineffective in the dense clumps that form from gravitational instabilities).  The feedback also generates a galactic fountain, with star forming regions ejecting gas out of the plane, which then falls back onto inactive regions.

Finally, we ask if cosmological simulations adopting the C-type of star formation algorithm can realistically be used to model star-forming galactic disks.  The answer appears to be a tentative yes, assuming that sufficiently high resolution is used to prevent spurious fragmentation in  the stable part of the disk.  As long as the correct parameters are chosen, we reproduce most of the features of the more realistic D-type model, in which stars form only in dense molecular clouds.  The exception is the fraction of stars formed in bound clumps, which is lower in the C models than the D models.  Models with supernova feedback reduce this discrepancy. 

Although throughout this paper we have emphasised the similarities in the large-scale features between the C- and D-type models, the results are not identical. A more detailed study of the differences and their effects on the disk's structure and interstellar medium will be a topic of future work.


\section{Conclusions}

We have performed high-resolution adaptive mesh refinement simulations of an isolated galactic disk evolved for more than 1 Gyr.  We include many of the physical processes which must be important for the long-term evolution of the gas in spiral galaxies including cooling, shocks, self-gravity, star  formation and supernova feedback in a global three-dimensional model.   Our adaptive-mesh methodology allows us to resolve scales from 100 kpc down to 25 pc, the size of typical giant molecular clouds.  The physical model for the galactic disk is clearly oversimplified in a number of respects: it does not include magnetic fields, cosmic rays, chemistry and the cooling/heating model is incomplete.  Still, this represents a substantial improvement over previous work in a number of ways (see the introduction for a discussion of previous simulation work) and represents some of the most realistic global disk simulations ever performed.

We performed a number of simulations while varying the input physics.   This included runs with cooling down to two minimum temperatures ($10^4$ K and 300 K), but no star formation.  A series of runs were performed with cooling and two different prescriptions for star formation, the first a cosmological-simulation inspired star formation algorithm  which allowed star formation at relatively low densities but with a low efficiency, and the second a more physically-minded algorithm which adopted a high density threshold (comparable to that found in giant molecular clouds) before stars could form.  These two forms we have denoted C-type and D-type, respectively.  We also performed some runs with spatial resolution two times better and mass resolution eight times better in order to investigate numerical convergence.  Finally, feedback from supernovae from massive stars was introduced.

Our results are summarized below:
\begin{itemize}

\item Gravitational instabilities grow as long as the Toomre $Q$ parameter is less than a critical value (0.6).  Outside of this region, no stars form (although stars can be scattered into this region).  This appears to be a well-resolved result, and does not depend on the star formation algorithm.

\item If no star formation occurs, the clumps merge and form more massive, denser clouds.  If star formation is permitted, stars form preferentially in the densest part of the clumps (particular for the D-type star formation algorithm).  Without some form of feedback, the clouds  are long-lived and convert a high fraction of their mass into stars.

\item Both star formation algorithms reproduce the slope of the observed relation between star formation and gas surface density.  This appears to be because clump formation is controlled by the dynamical time.  The C-type (cosmological) method can be tuned (with a sufficiently low efficiency parameter  $\epsilon < 0.005$) to reproduce the observed normalization of the relation as well.  The D-type method (with a high density threshold) produces too many stars and will require some additional form of feedback to match the normalization.  Energy input from type II supernovae does indeed decrease the star formation rate (although more feedback, such as photoionization, seems to be required to match observations).

\item A multiphase ISM is naturally reproduced with most of the mass ($>$ 80\%) in cold, dense clouds and peaks in the volume distribution at temperatures of approximately $10^4$ K and $10^{6.5}$ K.

\item Feedback from type II supernovae drives material out of the plane of the disk (which then falls back).  However, it does not increase the mean pressure in the plane of disk or generate large amounts of hot gas, or substantially increase the vertical scale-height of the gas.

\end{itemize}


We thank Adrianne Slyz, Julien Devriendt, Yuexing Li, Mordecai Mac Low and Frazer Pearce for useful discussions.  EJT and GLB acknowledge support from PPARC and GLB the Leverhulme Trust. Some simulations used in this paper were performed at the National Center for Supercomputing Applications.


\end{document}